\documentclass[11pt,a4paper]{article}
\usepackage[T1]{fontenc}
\usepackage[utf8]{inputenc}
\usepackage{lmodern}
\usepackage[margin=1in]{geometry}
\usepackage{amsmath,amssymb,amsthm,mathtools,bm}
\usepackage{microtype,booktabs,array,enumitem,needspace,etoolbox,graphicx,placeins}
\usepackage[round,authoryear]{natbib}
\usepackage{setspace}
\usepackage[colorlinks=true,linkcolor=blue!45!black,citecolor=blue!45!black,urlcolor=blue!45!black]{hyperref}
\usepackage{xcolor}
\hypersetup{pdftitle={Testing Treatment Effect Heterogeneity with Functional Outcomes},pdfauthor={Yongzhen Feng, Jie Li, Xiaojun Song}}
\setlist[enumerate]{leftmargin=*,itemsep=0.35em}
\newtheorem{assumption}{Assumption}[section]
\newtheorem{theorem}{Theorem}[section]
\newtheorem{proposition}[theorem]{Proposition}

\theoremstyle{remark}
\newtheorem{remark}[theorem]{Remark}
\BeforeBeginEnvironment{assumption}{\Needspace{4\baselineskip}}
\BeforeBeginEnvironment{theorem}{\Needspace{4\baselineskip}}
\BeforeBeginEnvironment{proposition}{\Needspace{4\baselineskip}}
\BeforeBeginEnvironment{lemma}{\Needspace{4\baselineskip}}
\pretocmd{\section}{\Needspace{5\baselineskip}}{}{}
\pretocmd{\subsection}{\Needspace{4\baselineskip}}{}{}
\newcommand{\HH}{\mathcal H}
\newcommand{\E}{\mathbb E}
\newcommand{\Prob}{\mathbb P}

\newcommand{\ind}{\mathbf 1}

\newcommand{\ii}{\mathrm i}
\newcommand{\GZ}{\Gamma_Z}
\newcommand{\mug}{\mu_{\Gamma_Z}}
\newcommand{\TU}{\widehat T_n}
\newcommand{\ip}[2]{\left\langle #1,#2\right\rangle}

\DeclareMathOperator{\Law}{Law}
\DeclareMathOperator{\Var}{Var}
\DeclareMathOperator{\tr}{tr}

\DeclareMathOperator{\Rea}{Re}

\title{Testing Treatment Effect Heterogeneity\\with Functional Outcomes}
\author{Yongzhen Feng\thanks{School of Mathematics and Statistics, Beijing Technology and Business University, Beijing, China.}
\and Jie Li\thanks{School of Statistics, Renmin University of China, Beijing, China. E-mail: \texttt{lijie\_stat@ruc.edu.cn}.}
\and Xiaojun Song\thanks{Department of Business Statistics and Econometrics, Guanghua School of Management, Peking University, Beijing, China.}}
\date{}
\begin{document}
\maketitle
\begin{abstract}
Assessing treatment-effect heterogeneity is important in causal inference, particularly for functional outcomes where effects may vary over time and the common effect is unknown. We propose a nonparametric procedure that assesses treatment-effect homogeneity through a testable marginal implication. By comparing within-arm difference distributions, the construction removes the unknown common functional effect without estimating it. We measure departures between these distributions through an integrated squared discrepancy of their characteristic functionals. A full-support Gaussian integrating measure yields an explicit bounded-kernel representation and identifies equality of the difference laws without requiring moment conditions on the outcomes.  We construct an unbiased generalized U-statistic that accounts for dependence among overlapping differences, derive its null, fixed-alternative, and local-alternative limits, and develop a centered multiplier bootstrap for inference. The framework extends to multiple treatment arms with a shared control. For computation, a random-direction approximation combined with a nine-state recursion reduces the cost to linear order in the sample size for fixed numerical settings. Simulation studies assess the finite-sample performance of the proposed procedure relative to several benchmark methods.  An application to the Irish gas smart-metering trial illustrates its use for assessing heterogeneity in functional treatment responses.
\end{abstract}
\noindent\textit{Preprint version note.} This version contains the complete main article and concise proof outlines. Detailed technical proofs and additional simulation and empirical appendices are not included.
\clearpage

\section{Introduction}
\label{sec:intro}

Understanding whether and how treatment effects vary across individuals is a fundamental question in causal inference and policy evaluation. Although the average treatment effect summarizes the overall impact of an intervention, it may conceal substantial differences in individual responses. Assessing such heterogeneity is important for designing targeted interventions \citep{imai2013estimating}. When outcomes are recorded repeatedly, treatment responses may differ in timing, duration, and shape as well as magnitude. Reducing each trajectory to a scalar summary can obscure these differences, motivating methods that assess heterogeneity across the entire response curve.

Functional data analysis provides a framework for studying trajectories as random functions, as described by \citet{RS05} and \citet{Wang-review2016}. A motivating application is the Irish gas smart-metering Customer Behaviour Trial, which examined the effects of billing, consumption feedback, and pricing interventions on residential gas use \citep{CER2011}. Our application involves four treatment groups and a shared control group, with consumption recorded every half hour. Each household's treatment assignment remained fixed, but its response could evolve throughout the observation period. The question is whether each intervention adds the same unknown effect curve to every household's untreated consumption curve. Homogeneity therefore concerns agreement across households, not constancy of the effect over time.

Testing this hypothesis presents three related challenges. First, only one potential outcome trajectory is observed for each individual. Even under random assignment, the marginal potential-outcome distributions do not determine their joint distribution, so individual homogeneity cannot generally be identified from observed outcomes alone. Second, a homogeneous effect may be an unknown nonzero function. Equality of the original outcome distributions is then too restrictive, and ordinary pooled label permutation is not generally valid. Third, differences in functional distributions may extend beyond their means and covariances, requiring a comparison that retains the distributional structure of the trajectories.

Research on scalar outcomes addresses several related notions of treatment effect heterogeneity. \citet{crump2008nonparametric} develop nonparametric tests for variation in conditional average effects across covariate-defined subpopulations. \citet{hsu2019testing} study heterogeneity in regression discontinuity designs, and \citet{SantAnna2021} develops tests for duration outcomes. A complementary literature examines variation not explained by observed covariates. \citet{Ding2016} develop randomization inference for treatment effect variation and address the nuisance parameter associated with an unknown common effect. \citet{Chung2021} construct an asymptotically valid permutation test using a martingale transformation of the empirical process. Particularly relevant to our approach, \citet{Ramirez2023} uses characteristic functions of symmetrized scalar outcomes to remove the unknown common effect. This shift-removing idea provides a starting point for our functional formulation.

Causal inference with functional outcomes has also received increasing attention. \citet{ecker2024causal} develop estimation and simultaneous inference for a functional average treatment effect, and \citet{testa2026doubly} propose a doubly robust estimator of this estimand. These methods focus on average causal response curves rather than homogeneity of individual effects. For functional hypothesis testing, \citet{cuevas2004anova} study mean comparisons, while \citet{Paparoditis2016} consider tests for equality of mean functions or covariance operators. \citet{Bugni2021} propose permutation tests for equality of functional distributions, and \citet{Wynne2022} develop kernel two-sample tests on function spaces. Our objective differs from direct two-sample distribution comparison because treatment and control outcomes may differ by an unknown common functional shift even under homogeneity.

We address this problem by comparing within-arm difference distributions through their characteristic functionals, which equal the squared moduli of the original outcome characteristic functionals. A common deterministic effect disappears from these differences, so neither the common effect nor individual effects need to be estimated. Equality of the difference laws is a necessary implication of individual homogeneity, not an equivalent characterization. We assess this implication using a Cram\'er--von Mises type integrated squared discrepancy. Retaining all within-arm pairs avoids restricting the analysis to a single set of disjoint pairs, but repeated use of individuals creates dependence among the differences. Valid inference must therefore account for their construction from the original independent samples.

The contributions of this paper are fourfold. First, we formulate the discrepancy on a real separable Hilbert space and show that, under a full-support Gaussian integrating measure, it vanishes exactly when the within-arm difference laws coincide. The Gaussian choice also yields an explicit bounded-kernel representation and a maximum mean discrepancy interpretation \citep{Gretton2012}, without requiring outcome moments. Second, we construct an unbiased generalized U-statistic and derive a projection expansion that accounts for overlapping differences. We establish null, fixed-alternative, and local-alternative limits and develop a centered multiplier bootstrap. The test has pointwise asymptotic level $\alpha$ under the observable null when the projection covariance is nonzero and is consistent against fixed alternatives with unequal difference laws. Third, we extend the procedure to multiple treatment arms, allowing intervention-specific common effect functions and accounting for dependence induced by a shared control. Fourth, we develop a random-direction implementation with a nine-state recursion that evaluates the distinct-index sums exactly at each sampled direction. For fixed numerical settings, both the finite-direction statistic and its multiplier critical value can be computed in time linear in sample size.

Simulation studies assess finite-sample calibration and power in single- and multiple-treatment settings. They demonstrate sensitivity to covariance-preserving distributional alternatives while revealing limitations near degeneracy and under extreme heavy tails. An application to monthly consumption in the Irish gas trial illustrates joint inference across interventions. The analysis finds no multiplicity-adjusted evidence against the tested restrictions during June--December 2010.

The remainder of the paper is organized as follows. Section~\ref{sec:setup} formulates the hypotheses and constructs the discrepancy. Section~\ref{sec:inference} develops estimation and inference, and Section~\ref{sec:multi} extends the method to multiple treatments. Section~\ref{sec:computation} presents the implementation. Sections~\ref{sec:simulation} and~\ref{sec:realdata} report the simulation studies and empirical application, respectively. Section~\ref{sec:discussion} concludes. Proofs and supplementary analyses are provided in the appendices.

\section{Hypotheses and a characteristic-functional discrepancy}
\label{sec:setup}

\subsection{Functional outcomes and the homogeneity hypothesis}
\label{sec:hypotheses}

Let $Y_i^0$ and $Y_i^1$ denote the potential outcome curves of individual $i$ under control and treatment, respectively, taking values in $L^2[0,1]$. Let $D_i\in\{0,1\}$ be the treatment indicator. The observed outcome is $Y_i=D_iY_i^1+(1-D_i)Y_i^0$, and the available data are $\{(D_i,Y_i)\}_{i=1}^n$. For each individual, only the potential outcome corresponding to the assigned treatment is observed. The other potential outcome is unavailable, so the individual treatment effect $Y_i^1-Y_i^0$ cannot be observed directly.

We assume that the potential-outcome pairs $\{(Y_i^0,Y_i^1)\}_{i=1}^n$ are independent and identically distributed across individuals and independent of the treatment assignment vector. It is worth noting that no restriction is imposed on the dependence between the two potential outcomes of the same individual. Let $F_d$ denote the Borel probability law of $Y_i^d$ on $L^2[0,1]$, and write $X_{d1},\ldots,X_{dn_d}$ for the observations in group $d$, where $n=n_0+n_1$. Conditional on the treatment assignment, these form two independent samples from $F_0$ and $F_1$, respectively. The subsequent analysis is conditional on the treatment assignment.

For a generic potential-outcome pair $(Y^0,Y^1)$, we consider the null hypothesis
\begin{equation*}
H_0^{\mathrm{hom}}\colon\quad Y^1-Y^0=\tau\ \text{almost surely for some deterministic }\tau\in L^2[0,1],
\end{equation*}
against $H_1^{\mathrm{hom}}$, under which $\Prob(Y^1-Y^0\ne\tau)>0$ for every $\tau\in L^2[0,1]$. The unknown function $\tau$ represents a treatment effect that is common to all individuals and may vary over time.

Under $H_0^{\mathrm{hom}}$, the marginal distributions satisfy $F_1=\Law(Y^0+\tau)$. Denote this restriction by $H_0^{\mathrm{shift}}$. Let $Y^{d\prime}$ be an independent copy of $Y^d$, and define $F_d^\Delta=\Law(Y^d-Y^{d\prime})$ for $d=0,1$. Under $H_0^{\mathrm{shift}}$, the difference between two independent draws from $F_1$ has the same distribution as the difference between two independent draws from $F_0$, since
\[
Y^1-Y^{1\prime}\overset{\mathrm d}{=}(Y^0+\tau)-(Y^{0\prime}+\tau)=Y^0-Y^{0\prime}.
\]
Thus $F_1^\Delta=F_0^\Delta$ is a consequence of homogeneous individual effects that does not involve the unknown function $\tau$. Moreover, each $F_d^\Delta$ is determined by a single marginal outcome distribution and can be estimated from observations within group $d$.

We therefore consider the observable testing problem
\begin{equation*}
H_0^{\mathrm{diff}}\colon F_1^\Delta=F_0^\Delta
\qquad\text{versus}\qquad
H_1^{\mathrm{diff}}\colon F_1^\Delta\ne F_0^\Delta.
\end{equation*}
Under the stated sampling model, rejection of $H_0^{\mathrm{diff}}$ provides evidence against $H_0^{\mathrm{hom}}$. Heterogeneous individual effects may nevertheless yield identical within-group difference distributions, so nonrejection does not establish homogeneity.

\subsection{Characteristic-functional discrepancy and Gaussian integration}
\label{sec:discrepancy}

The construction below applies to outcome laws on a real separable Hilbert space $\HH$, with inner product $\ip{\cdot}{\cdot}_{\HH}$ and associated norm $\|\cdot\|_{\HH}$. In the functional setting of Section~\ref{sec:hypotheses}, $\HH=L^2[0,1]$ and $\ip{f}{g}_{\HH}=\int_0^1 f(t)g(t)\,\mathrm dt$. We retain the notation $F_d$ and $F_d^\Delta$ for the outcome and independent-copy difference laws on $\HH$.

For $z\in\HH$, define the outcome characteristic functional by $\phi_d(z)=\E\exp\{\ii\ip{z}{Y^d}_{\HH}\}$. Independence of $Y^d$ and $Y^{d\prime}$ gives
\[
\phi_d^\Delta(z)=\E\exp\left\{\ii\ip{z}{Y^d-Y^{d\prime}}_{\HH}\right\}
=\phi_d(z)\phi_d(-z)=|\phi_d(z)|^2.
\]
By uniqueness of characteristic functionals, $H_0^{\mathrm{diff}}$ holds if and only if $\phi_1^\Delta(z)=\phi_0^\Delta(z)$ for every $z\in\HH$. Let $\Delta_\phi=\phi_1^\Delta-\phi_0^\Delta$. We measure departures from this restriction by the integrated squared discrepancy
\begin{equation*}
T_\mu(F_1,F_0)=\int_{\HH}\left\{\phi_1^\Delta(z)-\phi_0^\Delta(z)\right\}^2\,\mathrm d\mu(z),
\end{equation*}
where $\mu$ is a Borel probability measure on $\HH$.

Under $H_0^{\mathrm{diff}}$, the discrepancy vanishes for any choice of $\mu$. To ensure the converse, we choose an integrating measure with full support on $\HH$. A centered Gaussian measure is particularly convenient because it also yields an explicit kernel representation of $T_\mu$. Accordingly, let $\mu=\mug=N_{\HH}(0,\GZ)$, where the covariance operator satisfies the following assumption.

\begin{assumption}
\label{ass:gaussian}
The covariance operator $\GZ$ is deterministic, injective and trace class, and does not depend on $n$.
\end{assumption}

Henceforth, write $T=T_\mu(F_1,F_0)=\|\Delta_\phi\|_{L^2(\HH,\mu)}^2$. The following proposition establishes the identifying property of the discrepancy.

\begin{proposition}
\label{prop:ident}
Under Assumption~\ref{ass:gaussian}, for any Borel probability laws $F_1$ and $F_0$ on $\HH$,
\[
T=0\quad\Longleftrightarrow\quad F_1^\Delta=F_0^\Delta.
\]
\end{proposition}

For subsequent construction, define $k(u)=\exp\{-\tfrac12\ip{\GZ u}{u}_{\HH}\}$ for $u\in\HH$. The Gaussian characteristic-functional identity gives $\int_{\HH}\cos\{\ip{z}{u}_{\HH}\}\,\mathrm d\mu(z)=k(u)$, from which $T$ can be expressed in terms of expectations of this bounded kernel. In particular, $T$ is the squared maximum mean discrepancy between $F_1^\Delta$ and $F_0^\Delta$ for the two-argument kernel $(x,y)\mapsto k(x-y)$. The proof of Proposition~\ref{prop:ident} is given in Appendix~\ref{app:main-proofs}.

\section{Test statistic and asymptotic properties}
\label{sec:inference}

\subsection{Test statistic and its asymptotic expansion}
\label{sec:projection}

Expanding the square in $T$ gives two within-group expectations and one between-group expectation. Each within-group expectation involves two independent differences from the same population, whereas the between-group expectation involves one difference from each population. We estimate these quantities by U-statistics formed from distinct observations.

Define the symmetric kernels
\[
\begin{aligned}
h_{\mathrm w}(x_1,x_2,x_3,x_4)
={}&\frac13\bigl\{k(x_1+x_2-x_3-x_4)+k(x_1+x_3-x_2-x_4)\\
&\hspace{2.8em}+k(x_1+x_4-x_2-x_3)\bigr\}
\end{aligned}
\]
and
\[
h_{\mathrm b}(x_1,x_2;y_1,y_2)
=\frac12\left\{k(x_1-x_2+y_1-y_2)+k(x_1-x_2-y_1+y_2)\right\}.
\]
For $n_0,n_1\ge4$, let
\[
U_d=\binom{n_d}{4}^{-1}\sum_{i_1<i_2<i_3<i_4}
h_{\mathrm w}(X_{di_1},X_{di_2},X_{di_3},X_{di_4}),
\qquad d=0,1,
\]
and
\begin{equation}
\label{eq:U10}
U_{10}=\left\{\binom{n_1}{2}\binom{n_0}{2}\right\}^{-1}
\sum_{i<j}\sum_{r<s}h_{\mathrm b}(X_{1i},X_{1j};X_{0r},X_{0s}).
\end{equation}
The proposed test statistic is
\begin{equation}
\label{eq:TU}
\TU=U_1+U_0-2U_{10}.
\end{equation}

\begin{remark}[Exact unbiasedness]
\label{rem:unbiased}
Independent copies give $\E U_d=\|\phi_d^\Delta\|_{L^2(\HH,\mu)}^2$ and $\E U_{10}=\ip{\phi_1^\Delta}{\phi_0^\Delta}_{L^2(\HH,\mu)}$, so $\E\TU=T$. The use of four distinct observations in $U_d$ ensures independence of the two differences. Squaring empirical pair averages instead retains overlapping-index contributions of order $n^{-1}$, which affect the limiting null distribution at its leading order. The unbiased statistic need not be nonnegative. It is also invariant to separate deterministic shifts of the two samples.
\end{remark}

We establish the asymptotic null distribution of $\TU$, investigate its behavior under fixed and local alternatives, and then develop a multiplier bootstrap approximation. The sampling conditions in Section~\ref{sec:hypotheses} are maintained throughout, with the following additional condition on the group sizes.

\begin{assumption}
\label{ass:sampling}
As $n=n_0+n_1\to\infty$, $n_d/n\to\pi_d\in(0,1)$ for $d=0,1$.
\end{assumption}

We first establish a projection expansion of $\TU$, which provides the basis for the asymptotic results under the null and alternatives. For $x,y\in\HH$, define the pair feature $\Phi(x,y)\in L^2(\HH,\mu)$ by $\Phi(x,y)(z)=\cos\{\ip{z}{x-y}_{\HH}\}$. Here $L^2(\HH,\mu)$ is the real space with inner product $\ip{f}{g}=\int_{\HH}fg\,\mathrm d\mu$ and associated norm $\|\cdot\|$. Unless otherwise indicated, inner products and norms of frequency functions below refer to this space.

For group $d$, define
\begin{equation}
\label{eq:psi}
\psi_d(y)=2\left[\E\{\Phi(y,Y^{d\prime})\}-\phi_d^\Delta\right],
\qquad d=0,1.
\end{equation}
Let
\begin{equation}
\label{eq:phihat}
\widehat\phi_d^\Delta=\binom{n_d}{2}^{-1}\sum_{i<j}\Phi(X_{di},X_{dj}),
\qquad
\bar\psi_d=\frac1{n_d}\sum_i\psi_d(X_{di}).
\end{equation}
The Hoeffding decomposition of the pair average gives
\begin{equation}
\label{eq:pairaverage}
\widehat\phi_d^\Delta-\phi_d^\Delta=\bar\psi_d+U_{d,n}^{(2)},
\end{equation}
where $U_{d,n}^{(2)}$ is the canonical second-order term. Its properties are summarized in Appendix~\ref{app:main-proofs}. Define the quadratic term
\begin{equation*}
Q_n=\sum_{d=0}^1\frac1{n_d(n_d-1)}
\sum_{i\ne j}\ip{\psi_d(X_{di})}{\psi_d(X_{dj})}
-2\ip{\bar\psi_1}{\bar\psi_0}.
\end{equation*}

\begin{proposition}[Projection expansion]
\label{prop:expansion}
Under the sampling conditions in Section~\ref{sec:hypotheses} and Assumptions~\ref{ass:gaussian} and~\ref{ass:sampling},
\begin{equation}
\label{eq:expansion}
\TU=T+2\ip{\Delta_\phi}{\bar\psi_1-\bar\psi_0}
+Q_n+2\ip{\Delta_\phi}{U_{1,n}^{(2)}-U_{0,n}^{(2)}}+R_n.
\end{equation}
\end{proposition}

A brief outline of the expansion and the remainder bounds is given in Appendix~\ref{app:main-proofs}.

\subsection{Asymptotic null distribution}
\label{sec:null}

Under $H_0^{\mathrm{diff}}$, $\Delta_\phi=0$, so Proposition~\ref{prop:expansion} gives $\TU=Q_n+R_n$, with $nR_n=o_{\Prob}(1)$ as summarized in Appendix~\ref{app:main-proofs}. The limiting null distribution is therefore determined by $Q_n$. Let $C_d=\E\{\psi_d(Y^d)\otimes\psi_d(Y^d)\}$ and $C=\pi_1^{-1}C_1+\pi_0^{-1}C_0$, where $(u\otimes v)f=\ip{v}{f}u$.

\begin{theorem}[Asymptotic null distribution]
\label{thm:null}
Under Assumptions~\ref{ass:gaussian} and~\ref{ass:sampling}, if $H_0^{\mathrm{diff}}$ holds, then
\begin{equation*}
n\TU\Rightarrow L_C:=\|G\|^2-\tr(C)
=\sum_{r\ge1}\lambda_r(Z_r^2-1),
\end{equation*}
where $G\sim N_{L^2(\HH,\mu)}(0,C)$, the $\lambda_r$ are the eigenvalues of $C$, and the $Z_r$ are independent standard normal variables. The series converges in mean square.
\end{theorem}

The centering by $\tr(C)$ results from the exclusion of repeated indices in $\TU$. Since the weights depend on the unknown outcome distributions, the limiting null distribution is not distribution-free. We approximate its critical values using the multiplier procedure in Section~\ref{sec:bootstrap}.

\subsection{Asymptotic power under fixed and local alternatives}
\label{sec:power}
\label{sec:fixed}
\label{sec:local}

We first consider fixed outcome laws $F_1$ and $F_0$ under the alternative
\begin{equation*}
H_1^{\mathrm{diff}}\colon F_1^\Delta\ne F_0^\Delta,
\qquad\text{or equivalently}\qquad
T=\|\Delta_\phi\|^2>0.
\end{equation*}
Here the outcome laws, and hence $\Delta_\phi$, do not depend on $n$. Define
\[
\sigma_{\mathrm{alt}}^2
=4\sum_{d=0}^1\pi_d^{-1}
\Var\left\{\ip{\Delta_\phi}{\psi_d(Y^d)}\right\}.
\]

\begin{theorem}[Fixed alternatives]
\label{thm:fixed}
Under Assumptions~\ref{ass:gaussian} and~\ref{ass:sampling} and fixed outcome laws $F_1,F_0$, $\TU\to T$ in probability. If $\sigma_{\mathrm{alt}}^2>0$, then
\begin{equation*}
\sqrt n(\TU-T)\Rightarrow N(0,\sigma_{\mathrm{alt}}^2).
\end{equation*}
\end{theorem}

Under $H_1^{\mathrm{diff}}$, Proposition~\ref{prop:ident} gives $T>0$, and hence $n\TU\to\infty$ in probability. Together with the boundedness in probability of the bootstrap critical value established in Section~\ref{sec:bootstrap}, this implies consistency against fixed alternatives. The positive-variance condition is needed for the nondegenerate normal limit, but not for consistency.

We next consider a sequence of local alternatives approaching $H_0^{\mathrm{diff}}$. For each $n$, let the treatment and control samples have laws $F_{1,n}$ and $F_{0,n}$, respectively, and define $\phi_{d,n}^\Delta$, $\psi_{d,n}$ and $C_{d,n}$ under these laws. The samples remain independent, with independent and identically distributed observations within each group. Writing $\Delta_{\phi,n}=\phi_{1,n}^\Delta-\phi_{0,n}^\Delta$, we consider
\begin{equation*}
H_{1,n}^{\mathrm{diff}}\colon\quad
\Delta_{\phi,n}=n^{-1/2}\delta_\phi+o(n^{-1/2})
\quad\text{in }L^2(\HH,\mu),
\end{equation*}
where $\delta_\phi$ is a fixed nonzero element of $L^2(\HH,\mu)$. The integrating measure remains fixed. Let $C_n=(n/n_1)C_{1,n}+(n/n_0)C_{0,n}$, and denote the trace norm by $\|\cdot\|_1$.

\begin{assumption}
\label{ass:local}
Along the local sequence, $\|C_n-C\|_1\to0$ for a nonnegative trace-class operator $C$ on $L^2(\HH,\mu)$.
\end{assumption}

\begin{theorem}[Local alternatives]
\label{thm:local}
Under Assumptions~\ref{ass:gaussian}, \ref{ass:sampling} and~\ref{ass:local} and $H_{1,n}^{\mathrm{diff}}$,
\begin{equation*}
n\TU\Rightarrow\|G+\delta_\phi\|^2-\tr(C),
\qquad G\sim N_{L^2(\HH,\mu)}(0,C).
\end{equation*}
The same conclusion holds for sequences satisfying $\sqrt n\,\Delta_{\phi,n}\to0$, with $\delta_\phi=0$.
\end{theorem}

The local alternatives are specified at rate $n^{-1/2}$ in the difference characteristic functional, corresponding to rate $n^{-1}$ in the population discrepancy. Theorem~\ref{thm:local}, together with the bootstrap approximation below, gives the limiting rejection probability under these alternatives.

\subsection{Multiplier bootstrap approximation}
\label{sec:bootstrap}

The limiting null distribution in Theorem~\ref{thm:null} depends on the unknown covariance operator $C$. We approximate it by applying independent multipliers to empirical versions of the projections $\psi_d$, rather than estimating the eigenvalues of $C$ individually.

Replacing the population expectations in $\psi_d$ by within-group averages gives
\begin{equation}
\label{eq:ahat}
\widehat\psi_{di}
=2\left\{\frac1{n_d-1}\sum_{j\ne i}\Phi(X_{di},X_{dj})
-\widehat\phi_d^\Delta\right\},
\qquad i=1,\ldots,n_d.
\end{equation}
These empirical projections satisfy $\sum_{i=1}^{n_d}\widehat\psi_{di}=0$. Define $\widehat C_d=n_d^{-1}\sum_{i=1}^{n_d}\widehat\psi_{di}\otimes\widehat\psi_{di}$ and $\widehat C_n=(n/n_1)\widehat C_1+(n/n_0)\widehat C_0$.

Let $\{W_{di}\}$ be independent and identically distributed multipliers, independent of the observations, with a fixed distribution satisfying $\E W=0$, $\E W^2=1$ and $\E W^4<\infty$. Define
\begin{equation*}
G_n^*=\sqrt n\left\{
\frac1{n_1}\sum_{i=1}^{n_1}W_{1i}\widehat\psi_{1i}
-\frac1{n_0}\sum_{j=1}^{n_0}W_{0j}\widehat\psi_{0j}
\right\}.
\end{equation*}
Conditional on the observations, $G_n^*$ has mean zero and covariance operator $\widehat C_n$. We therefore use the centered quadratic statistic
\begin{equation}
\label{eq:bootstrap}
L_n^*=\|G_n^*\|^2-\tr(\widehat C_n)
\end{equation}
to approximate $L_C$. Let $c_{1-\alpha}^*$ be its conditional $(1-\alpha)$ quantile. At significance level $\alpha\in(0,1)$, the test rejects $H_0^{\mathrm{diff}}$ when $n\TU>c_{1-\alpha}^*$.

\begin{theorem}[Bootstrap validity and consistency]
\label{thm:bootstrap}
Under Assumptions~\ref{ass:gaussian} and~\ref{ass:sampling} and $H_0^{\mathrm{diff}}$, the conditional law of $L_n^*$ converges weakly in probability to the law of $L_C$. If $C\ne0$, then
\[
\Prob\left\{n\TU>c_{1-\alpha}^*\right\}\longrightarrow\alpha.
\]
Under fixed outcome laws satisfying $F_1^\Delta\ne F_0^\Delta$, the rejection probability converges to one.

Under Assumptions~\ref{ass:gaussian}, \ref{ass:sampling} and~\ref{ass:local}, along the local sequence in Section~\ref{sec:power}, the conditional law of $L_n^*$ converges weakly in probability to the law of $L_C$, where $C$ is the covariance limit of that sequence. If $C\ne0$, then
\begin{equation}
\label{eq:localpower}
\Prob\left\{n\TU>c_{1-\alpha}^*\right\}
\longrightarrow
\Prob\left\{\|G+\delta_\phi\|^2-\tr(C)>c_{1-\alpha}\right\},
\end{equation}
where $c_{1-\alpha}$ is the $(1-\alpha)$ quantile of $L_C$.
\end{theorem}

The theorem establishes pointwise asymptotic validity under the observable null and consistency against fixed alternatives. The distributional approximation remains valid when $C=0$, but the asymptotic level and local rejection-probability statements require $C\ne0$. In practice, the conditional quantile is approximated using repeated multiplier draws. Details of the implementation are given in Section~\ref{sec:computation}.

\subsection{Multiple treatments with a shared control}
\label{sec:multi}

Suppose that there are a fixed number $m$ of active treatment groups and a control group indexed by $0$. For $d=0,\ldots,m$, let $X_{d1},\ldots,X_{dn_d}$ be an independent and identically distributed sample from $F_d$, with independence across groups. Write $n=\sum_{d=0}^m n_d$, and assume $n_d\ge4$ and $n_d/n\to\pi_d>0$ for every group. Define $F_d^\Delta$, $\phi_d^\Delta$, $\psi_d$ and $C_d$ as in the two-group case, using the same integrating measure satisfying Assumption~\ref{ass:gaussian}.

The joint null requires $F_d^\Delta=F_0^\Delta$ for all $d=1,\ldots,m$, against the alternative that at least one equality fails. It is a necessary consequence of homogeneous individual effects within each treatment, allowing a different deterministic effect $\tau_d$ for each treatment. With $\Delta_{\phi,d}=\phi_d^\Delta-\phi_0^\Delta$, define
\[
T_m=\sum_{d=1}^m\|\Delta_{\phi,d}\|^2,
\qquad
\widehat T_{m,n}=\sum_{d=1}^m(U_d+U_0-2U_{d0}),
\]
where $U_{d0}$ is the between-group statistic~\eqref{eq:U10} with treatment group $1$ replaced by group $d$. Proposition~\ref{prop:ident} implies that $T_m=0$ if and only if the joint null holds.

Although observations from different groups are independent, the treatment--control comparisons are dependent through the shared control sample. For the joint Gaussian limit, define the covariance operator $\boldsymbol C_m$ on $\{L^2(\HH,\mu)\}^m$, equipped with the sum inner product, by its blocks
\[
(\boldsymbol C_m)_{d\ell}
=\ind\{d=\ell\}\pi_d^{-1}C_d+\pi_0^{-1}C_0,
\qquad d,\ell=1,\ldots,m.
\]
The shared-control contribution $\pi_0^{-1}C_0$ appears in every block, including those corresponding to distinct comparisons.

The bootstrap reproduces this dependence by reusing the same control multipliers across all comparisons. For $d=1,\ldots,m$, define
\begin{equation}
\label{eq:multiGstar}
G_{d,n}^*=\sqrt n\left\{
\frac1{n_d}\sum_{i=1}^{n_d}W_{di}\widehat\psi_{di}
-\frac1{n_0}\sum_{i=1}^{n_0}W_{0i}\widehat\psi_{0i}
\right\},
\end{equation}
where the multipliers satisfy the conditions in Section~\ref{sec:bootstrap}. A single set $\{W_{0i}\}_{i=1}^{n_0}$ is used for all treatment--control comparisons, and the treatment-group multipliers are independent across groups. Let $\boldsymbol G_{m,n}^*=(G_{1,n}^*,\ldots,G_{m,n}^*)$. Its conditional covariance operator $\widehat{\boldsymbol C}_{m,n}$ has blocks
\[
(\widehat{\boldsymbol C}_{m,n})_{d\ell}
=\ind\{d=\ell\}\frac n{n_d}\widehat C_d
+\frac n{n_0}\widehat C_0.
\]
The joint bootstrap statistic is $L_{m,n}^*=\|\boldsymbol G_{m,n}^*\|^2-\tr(\widehat{\boldsymbol C}_{m,n})$.

\begin{theorem}[Multiple treatments]
\label{thm:multi}
Under the preceding sampling conditions and Assumption~\ref{ass:gaussian}, if $F_d^\Delta=F_0^\Delta$ for all $d=1,\ldots,m$, then
\begin{equation*}
n\widehat T_{m,n}\Rightarrow
\|\boldsymbol G_m\|^2-\tr(\boldsymbol C_m),
\end{equation*}
where $\boldsymbol G_m$ is a centered Gaussian element of $\{L^2(\HH,\mu)\}^m$ with covariance operator $\boldsymbol C_m$. The conditional law of $L_{m,n}^*$ converges weakly in probability to this limiting law. If $\boldsymbol C_m\ne0$, the test based on its conditional $(1-\alpha)$ quantile has asymptotic level $\alpha$. Under fixed alternatives with $T_m>0$, its rejection probability converges to one.
\end{theorem}

A proof outline for Theorem~\ref{thm:multi} is given in Appendix~\ref{app:main-proofs}. The fixed- and local-alternative extensions follow from the corresponding two-group results.

\section{Implementation}
\label{sec:computation}

Direct evaluation of \eqref{eq:TU} requires summing over within-arm quadruples and combinations of two treated and two control observations. The resulting computational cost is $O\left(n^4\right)$ when the allocation proportions are bounded away from zero. We avoid this enumeration by approximating the Gaussian integral with random directions and evaluating the distinct-index sums recursively. The resulting calculations provide both the test statistic and the empirical projections needed for multiplier inference.

\subsection{Gaussian directions}
\label{sec:computation-directions}

Let $(e_k)$ be an orthonormal eigenbasis of $\GZ$, with corresponding eigenvalues $(\gamma_k)$. We retain $K$ basis coefficients per curve and approximate the Gaussian integral using $M$ independent random directions. Following the random-feature approach of \citet{Rahimi2007}, we generate
\[
Z_m^{\left(K\right)}
=
\sum_{k=1}^{K}\sqrt{\gamma_k}\,\xi_{mk}e_k,
\qquad m=1,\ldots,M,
\]
where the $\xi_{mk}$ are independent standard normal variables generated independently of the observations. Inner products with $Z_m^{\left(K\right)}$ are computed from the stored basis scores. The same set of directions is used for the observed statistic and all bootstrap replications.

\begin{remark}[Curves observed on a grid]
\label{rem:reconstruction}
For grid observations, basis scores may be computed by quadrature, $\left\langle X_{di},e_k\right\rangle_{\HH} \approx\sum_\ell w_\ell X_{di}\left(t_\ell\right)e_k\left(t_\ell\right)$, where $w_\ell$ are quadrature weights, or from curves reconstructed using a prespecified smoother. The same basis and reconstruction rule are used across arms. A common linear reconstruction preserves deterministic additive shifts. With noisy or irregular observations, preservation of the distributional restriction also depends on the noise and observation mechanisms. The test concerns the curves supplied to the algorithm. Interpretation for latent trajectories requires additional reconstruction conditions, particularly with sparse observations.
\end{remark}

\subsection{Nine-state recursion}
\label{sec:computation-recursion}

For each arm $d=0,1$ and a fixed direction $z$, write $q_{di}=\exp\left\{\ii\left\langle z,X_{di}\right\rangle_{\HH}\right\}$ and $S_d=\sum_{i=1}^{n_d}q_{di}$. The pair-based estimator in \eqref{eq:phihat} can then be evaluated as
\begin{equation}
\label{eq:ghatnumeric}
\widehat\phi_d^\Delta\left(z\right)
=
\frac{\left|S_d\right|^2-n_d}{n_d\left(n_d-1\right)}.
\end{equation}
An unbiased estimator of $\left\{\phi_d^\Delta\left(z\right)\right\}^2$ requires products involving four distinct observations. Directly squaring $\widehat\phi_d^\Delta\left(z\right)$ would introduce repeated indices. We therefore compute the required fourth-order sums separately.

Motivated by dynamic-programming approaches to computing higher-order U-statistics \citep{Zhang2025}, we derive a nine-state recursion for these distinct-index sums. Let $P_{rs}$ accumulate products containing $r$ nonconjugate factors and $s$ conjugate factors, with each factor drawn from a different observation. Only $r,s\in\left\{0,1,2\right\}$ are needed, so the calculation uses a $3\times3$ array. Initialize $P_{00}=1$ and all other entries at zero. For each observation, update
\begin{equation}
\label{eq:recursion}
P_{rs}^{\mathrm{new}}
=
P_{rs}^{\mathrm{old}}
+
q_{di}P_{r-1,s}^{\mathrm{old}}
+
\overline{q_{di}}P_{r,s-1}^{\mathrm{old}},
\qquad r,s=0,1,2,
\end{equation}
where entries with a negative index are zero. The three terms correspond to omitting the observation or selecting it as a nonconjugate or conjugate factor. Every update uses the previous array, preventing an observation from entering the same product twice. After processing all observations in the arm, set
\begin{equation}
\label{eq:vhatnumeric}
\widehat v_d\left(z\right)
=
\frac{4\Rea P_{22}}{\left(n_d\right)_4},
\end{equation}
where $\left(n_d\right)_4=n_d\left(n_d-1\right)\left(n_d-2\right)\left(n_d-3\right)$. The factor $4=2!2!$ accounts for the possible orders of the two selected indices of each type. The resulting $\widehat v_d\left(z\right)$ is an unbiased estimator of $\left\{\phi_d^\Delta\left(z\right)\right\}^2$. The array is reset for each arm and direction. The coefficient interpretation follows from selecting at most one term from each factor of the generating polynomial.

Repeating these calculations for both arms at all directions gives
\[
\widehat T_{\mathrm{freq}}
=
\frac{1}{M}\sum_{m=1}^{M}
\left\{
\widehat v_1\left(Z_m^{\left(K\right)}\right)
+
\widehat v_0\left(Z_m^{\left(K\right)}\right)
-
2\widehat\phi_1^\Delta\left(Z_m^{\left(K\right)}\right)
 \widehat\phi_0^\Delta\left(Z_m^{\left(K\right)}\right)
\right\}.
\]
The cross-arm contribution is the product of the two pair-based estimates because the arm samples are independent. No additional quadruple summation is required. The recursion evaluates this finite-direction statistic exactly in algebraic terms. Approximation to the full Gaussian-measure statistic arises only from spectral truncation and the finite direction bank.

\subsection{Multiplier calibration}
\label{sec:computation-multiplier}

The empirical projections in \eqref{eq:ahat} are computed from the same quantities as
\begin{equation}
\label{eq:scoreRF}
\widehat\psi_{di}\left(z\right)
=
2\left\{
\frac{\Rea\left(q_{di}\overline{S_d}\right)-1}{n_d-1}
-
\widehat\phi_d^\Delta\left(z\right)
\right\}.
\end{equation}
These scores satisfy $\sum_{i=1}^{n_d}\widehat\psi_{di}\left(z\right)=0$. For each bootstrap replication, draw independent multipliers $W_{di}$ as specified in Section~\ref{sec:bootstrap} and calculate
\begin{equation}
\label{eq:rfbootstrap}
\begin{aligned}
L_{\mathrm{freq}}^*
={}&
\frac{n}{M}\sum_{m=1}^{M}
\left\{
\frac{1}{n_1}\sum_{i=1}^{n_1}
W_{1i}\widehat\psi_{1i}\left(Z_m^{\left(K\right)}\right)
-
\frac{1}{n_0}\sum_{j=1}^{n_0}
W_{0j}\widehat\psi_{0j}\left(Z_m^{\left(K\right)}\right)
\right\}^{2}\\
&-\frac{n}{M}\sum_{m=1}^{M}\sum_{d=0}^{1}
\frac{1}{n_d^2}\sum_{i=1}^{n_d}
\widehat\psi_{di}\left(Z_m^{\left(K\right)}\right)^2.
\end{aligned}
\end{equation}
The second term subtracts the conditional mean of the quadratic form, as in \eqref{eq:bootstrap}. Each individual receives the same multiplier at every direction, preserving dependence across directions. After $B$ replications, reject when $n\widehat T_{\mathrm{freq}}$ exceeds the empirical $\left(1-\alpha\right)$ quantile of the bootstrap values.

For fixed $K$ and $M$, this calibration applies to the finite-direction criterion. It is asymptotically valid under the stated conditions when the corresponding projection covariance is nonzero. Consistency requires a positive discrepancy on the sampled direction bank, rather than merely a positive discrepancy under the full Gaussian measure. This finite-bank condition is distinct from the full-support identification result in Proposition~\ref{prop:ident}.

The implementation extends to multiple treatments by computing the same armwise quantities and combining them as in Section~\ref{sec:multi}. The joint multiplier statistic uses the same control multipliers across all treatment--control contrasts, preserving the dependence induced by the shared control sample. The values of $K$, $M$, and $B$, together with the spectrum and random seeds, specify the numerical implementation.

\begin{remark}[Computational cost]
\label{rem:complexity}
Given $K$ stored basis scores per curve, direct evaluation of the Gaussian-kernel statistic costs $O\left(Kn^4\right)$. The random-direction implementation requires $O\left(MKn\right)$ operations to construct the phases, $O\left(Mn\right)$ to evaluate the statistic and projection scores, and $O\left(BMn\right)$ for the multiplier replications. The total cost is therefore $O\left\{Mn\left(K+B\right)\right\}$, which is linear in $n$ for fixed $K$, $M$, and $B$. This reduction comes from reorganizing the distinct-index sums and does not change the order of the underlying U-statistic.
\end{remark}

\Needspace{8\baselineskip}
\section{Simulation study}\label{sec:simulation}

\subsection{Null settings and benchmarks}
\label{sec:simdesign}

We  examine the finite-sample size of the proposed test under a range of outcome distributions and sampling configurations. Because the proposed procedure and the benchmark methods considered below are invariant to arm-specific deterministic shifts, the common treatment effect can be set to zero without loss of generality for size assessment. We therefore vary the outcome distribution, tail behavior, discreteness, allocation proportions, and degree of degeneracy across the null designs.

Let $\{e_j:j\geq1\}$ be the cosine basis of $L^2[0,1]$, with $e_1(t)=1$ and $e_j(t)=\sqrt{2}\cos\{(j-1)\pi t\}$ for $j\geq2$. The Gaussian reference process is
\begin{equation}
Y_i^0(t)
=
\sum_{j=1}^{100}j^{-0.55}\xi_{ij}e_j(t),
\label{eq:simbase}
\end{equation}
where the coefficients $\xi_{ij}$ are independent and identically distributed as $N(0,1)$. We consider total sample sizes $n=100,200$ in the single-treatment setting and $n=120,240$ in the multiple-treatment setting. The following eight null designs are considered in both settings, with observations independent across individuals and arms.
\begin{enumerate}[label=(N\arabic*),itemsep=0.4em]
\item \textit{Gaussian.} Every arm follows the Gaussian reference law in~\eqref{eq:simbase}.

\item \textit{Skewed.} Replace each $\xi_{ij}$ by $E_{ij}-1$, where $E_{ij}\sim\operatorname{Exp}(1)$.

\item \textit{Heavy-tailed with finite variance.} Replace each $\xi_{ij}$ by $T_{ij}/\sqrt{3}$, where $T_{ij}$ follows a Student $t$ distribution with three degrees of freedom. The resulting scores have unit variance and infinite fourth moment.

\item \textit{Discrete.} Replace each $\xi_{ij}$ by $(B_{ij}-0.2)/0.4$, where $B_{ij}\sim\operatorname{Bernoulli}(0.2)$.

\item \textit{Unequal allocation.} Use the Gaussian law with $(n_0,n_1)=(3n/5,2n/5)$ in the single-treatment setting and $(n_0,n_1,n_2,n_3)=(n/2,n/6,n/6,n/6)$ in the multiple-treatment setting.

\item \textit{Difference-law null without a marginal shift.} Independently draw $A\sim\operatorname{Bernoulli}(0.2)$ and $B\sim\operatorname{Bernoulli}(0.3)$. The control law is $\Law\{(A+2B)e_1\}$ and each treatment law is $\Law\{(A-2B)e_1\}$. These distributions satisfy $H_0^{\mathrm{diff}}$ but not $H_0^{\mathrm{shift}}$. These claims follow by comparing the characteristic functions and support probabilities of the two laws.

\item \textit{Near-degenerate projection.} Every arm follows $\Law(Be_1)$ with $B\sim\operatorname{Bernoulli}(0.48)$. The first-order covariance is positive but small and vanishes when the success probability equals $1/2$.

\item \textit{No finite first moment.} Replace each $\xi_{ij}$ by an independent standard Cauchy variable. Each curve remains almost surely finite in the 100-dimensional subspace.
\end{enumerate}

Except for (N6), all designs satisfy the homogeneity null with a zero common effect. Design (N6) satisfies only the weaker observable difference-law null. We compare the proposed procedure with the following benchmark methods.
\begin{enumerate}
\item[(i)] \textit{Covariance-operator method (COVM).} Let $\bar X_d=n_d^{-1}\sum_iX_{di}$ and $\widehat\Sigma_d=n_d^{-1}\sum_i(X_{di}-\bar X_d)\otimes(X_{di}-\bar X_d)$. For $m$ treatment arms, define
\[
S_{\mathrm{COVM}}
=
n\sum_{a=1}^{m}
\left\|
\widehat\Sigma_a-\widehat\Sigma_0
\right\|_{\mathrm{HS}}^2.
\]
This benchmark compares covariance structure across groups and is closely related to functional covariance-operator tests such as \citet{Paparoditis2016}.

\item[(ii)] \textit{Symmetry-preserving maximum mean discrepancy test (S-MMD).} Within each arm, form disjoint differences $D_{dr}=X_{d,2r-1}-X_{d,2r}$ for $r=1,\ldots,r_d$, where $r_d=\lfloor n_d/2\rfloor$, and represent each difference by $(\delta_D+\delta_{-D})/2$. Let $\nu_a=r_a^{-1}\sum_r\delta_{D_{ar}}-r_0^{-1}\sum_r\delta_{D_{0r}}$ and $k_{\pm}(u,v)=\{k(u-v)+k(u+v)\}/2$. We use
\[
S_{\mathrm{S\text{-}MMD}}
=
\sum_{a=1}^{m}
\iint k_{\pm}(u,v)\,
\mathrm d\nu_a(u)\,
\mathrm d\nu_a(v).
\]
This is a symmetry-preserving version of the maximum mean discrepancy criterion of \citet{Gretton2012} applied to independent within-arm differences.

\item[(iii)] \textit{Symmetry-preserving energy-distance test (S-ED).} Using the same paired differences, let $d_{\pm}(u,v)=\{\|u-v\|_{\HH}+\|u+v\|_{\HH}\}/2$ and define
\[
S_{\mathrm{S\text{-}ED}}
=
-
\sum_{a=1}^{m}
\iint d_{\pm}(u,v)\,
\mathrm d\nu_a(u)\,
\mathrm d\nu_a(v).
\]
This benchmark applies energy distance to the symmetrized difference distributions and, under finite first moments, identifies equality of probability laws on separable Hilbert spaces \citep{Lyons2013,Sejdinovic2013}.
\end{enumerate}

For calibration, the proposed procedure and COVM use Rademacher multipliers, whereas S-MMD and S-ED use label permutations of the independent paired differences. The latter two methods use the same disjoint pairs and the same permutations, with $\{D,-D\}$ treated as a single resampling unit.

For all simulations, we retain $K=100$ coefficients and use $M=256$ fixed Gaussian directions with spectrum $\gamma_j=2j^{-1.2}$. The direction bank is generated independently of the outcomes. Each method uses $B=999$ resamples, with $\widehat p=\{1+\sum_{b=1}^{999}\ind(S_b^*\geq S_{\rm obs})\}/1000$. Each rejection frequency is based on 500 independently generated data sets, with all methods evaluated on the same data within each replication. Rejection frequencies are reported at the nominal 0.05 level.

\subsection{A single treatment against control}
\label{sec:simsingle}

We first consider the setting with a single treatment group and a control group. For size assessment, we use the eight null designs introduced in Section~\ref{sec:simdesign}. For the power experiments, the untreated potential outcome $Y_i^0$ follows the Gaussian reference law in~\eqref{eq:simbase}. We keep this baseline law fixed across all alternatives and vary the individual treatment effect $\tau_i$, allowing it to depend on $Y_i^0$. Under treatment,
\[
Y_i^1(t)=Y_i^0(t)+\tau_i(t).
\]
The treatment and control samples consist of independent individuals drawn from the same baseline distribution. Let $\Phi_{\mathrm N}$ denote the standard normal distribution function. We consider the following six alternatives.

\begin{enumerate}[label=(\roman*),itemsep=0.45em]

\item \textit{Baseline-linked sinusoidal amplitudes.} Set $\tau_i(t)=2\Phi_{\mathrm N}(\xi_{i1})\sin(2\pi t)$. The amplitude is marginally uniform on $(0,2)$ and depends on the first baseline score.

\item \textit{Three effect levels.} Set $\tau_i(t)=L_i$, where $\Prob(L_i=0)=\Prob(L_i=1)=\Prob(L_i=2)=1/3$.

\item \textit{Independent sinusoidal amplitudes.} Set $\tau_i(t)=4U_i\sin(2\pi t)$, where $U_i\stackrel{\mathrm{i.i.d.}}{\sim}\operatorname{Unif}(0,1)$.

\item \textit{Mixed linear and nonlinear effects.} Let $G_i\stackrel{\mathrm{i.i.d.}}{\sim}\operatorname{Bernoulli}(1/2)$ and set
\[
\tau_i(t)=
\begin{cases}
t, & G_i=1,\\
2\sin(2\pi t), & G_i=0.
\end{cases}
\]

\item \textit{Piecewise heterogeneous effects.} Let $U_i\stackrel{\mathrm{i.i.d.}}{\sim}\operatorname{Unif}(0,1)$ and $b_i=3U_i$, and set
\[
\tau_i(t)=
\begin{cases}
0, & 0\le t\le0.3,\\
b_i, & 0.3<t\le0.7,\\
10-b_i, & 0.7<t\le1.
\end{cases}
\]
The common component after $t=0.7$ cancels from within-arm differences, while the component involving $b_i$ generates heterogeneity.

\item \textit{Covariance-matched higher-order alternative.} Let $Q_i=(\xi_{i1}^2-1)/\sqrt{2}$ and set $\tau_i(t)=(Q_i-\xi_{i1})e_1(t)$. Thus the first basis coefficient changes from $\xi_{i1}\sim N(0,1)$ under control to $Q_i$ under treatment, while all remaining coefficients are unchanged. Since $\E Q_i=0$ and $\Var(Q_i)=1$, the two potential-outcome laws have the same mean and covariance operator, $\operatorname{Cov}(Y_i^1)=\operatorname{Cov}(Y_i^0)=\sum_{j=1}^{100}j^{-1.1}e_j\otimes e_j$. Their difference distributions nevertheless differ because $\phi_0^\Delta(se_1)=e^{-s^2}$ whereas $\phi_1^\Delta(se_1)=(1+2s^2)^{-1/2}$. This design isolates higher-order heterogeneity that is not reflected in covariance differences.

\end{enumerate}

In the three effect-level, independent sinusoidal, mixed-shape, and piecewise-effect designs, the variables determining $\tau_i$ are independent of $Y_i^0$. The baseline-linked sinusoidal and covariance-matched alternatives instead allow the treatment effect to depend on the baseline process.

Table~\ref{tab:simsize} shows that the proposed test maintains reasonably stable size across a broad range of null distributions and sampling configurations. Its rejection frequencies remain close to the nominal level under the Gaussian, skewed, Student $t_3$, discrete, unequal-allocation, and weak difference-law nulls. In particular, the frequencies of 0.046 and 0.054 under the weak difference-law null confirm that calibration does not rely on the treatment and control distributions being related by a deterministic shift. The symmetry-preserving paired benchmarks also exhibit stable calibration in these settings. The more extreme designs reveal finite-sample limitations of the proposed procedure, with overrejection under the near-degenerate projection design and strong conservativeness under the Cauchy design. For COVM, the multiplier justification does not cover the Student $t_3$ setting because the fourth moment is infinite, and the method is not reported for the Cauchy setting because the population covariance is undefined.

\begin{table}[!htbp]
\centering
\caption{Null rejection frequencies for one treatment against control}
\label{tab:simsize}
\scriptsize
\setlength{\tabcolsep}{3pt}
\begin{tabular}{lrrrrrrrr}
\toprule
& \multicolumn{4}{c}{$n=100$} & \multicolumn{4}{c}{$n=200$}\\
\cmidrule(lr){2-5}\cmidrule(lr){6-9}
Design
& Proposed & COVM & S-MMD & S-ED
& Proposed & COVM & S-MMD & S-ED\\
\midrule
(N1) Gaussian
& 0.054 & 0.082 & 0.034 & 0.040
& 0.056 & 0.060 & 0.048 & 0.062\\
(N2) Skewed
& 0.070 & 0.094 & 0.052 & 0.046
& 0.040 & 0.052 & 0.042 & 0.052\\
(N3) Student $t_3$
& 0.058 & 0.044 & 0.048 & 0.068
& 0.058 & 0.050 & 0.042 & 0.044\\
(N4) Discrete
& 0.058 & 0.078 & 0.052 & 0.050
& 0.072 & 0.062 & 0.058 & 0.060\\
(N5) Unequal allocation
& 0.056 & 0.086 & 0.056 & 0.052
& 0.058 & 0.072 & 0.052 & 0.052\\
(N6) Weak null
& 0.046 & 0.060 & 0.038 & 0.028
& 0.054 & 0.052 & 0.064 & 0.058\\
(N7) Near-degenerate
& 0.110 & 0.000 & 0.036 & 0.036
& 0.122 & 0.000 & 0.030 & 0.030\\
(N8) Cauchy
& 0.002 & --- & 0.046 & 0.054
& 0.000 & --- & 0.042 & 0.044\\
\bottomrule
\end{tabular}
\end{table}

Table~\ref{tab:simpower} shows that the proposed test gains substantial power as the sample size increases and consistently outperforms the two symmetry-preserving paired benchmarks across all six alternatives. COVM can have higher rejection frequencies when heterogeneity changes the covariance structure, although these comparisons should be interpreted together with its finite-sample size behavior. The covariance-matched higher-order alternative provides a particularly informative contrast. When the treatment and control covariance operators are identical, the proposed rejection frequency increases from 0.664 to 0.926, while COVM remains close to its own null rejection level and the paired benchmarks are substantially less sensitive. Overall, the results demonstrate that the proposed procedure is sensitive to a broad range of treatment-effect heterogeneity, including distributional differences that are not reflected in covariance.

\begin{table}[!htbp]
\centering
\caption{Alternative rejection frequencies for one treatment against control}
\label{tab:simpower}
\scriptsize
\setlength{\tabcolsep}{3pt}
\begin{tabular}{lrrrrrrrr}
\toprule
& \multicolumn{4}{c}{$n=100$} & \multicolumn{4}{c}{$n=200$}\\
\cmidrule(lr){2-5}\cmidrule(lr){6-9}
Design
& Proposed & COVM & S-MMD & S-ED
& Proposed & COVM & S-MMD & S-ED\\
\midrule
(i) Baseline-linked sinusoid
& 0.444 & 0.582 & 0.122 & 0.180
& 0.854 & 0.894 & 0.210 & 0.328\\
(ii) Three effect levels
& 0.396 & 0.452 & 0.110 & 0.190
& 0.636 & 0.706 & 0.200 & 0.358\\
(iii) Independent sinusoid
& 0.532 & 0.698 & 0.170 & 0.316
& 0.870 & 0.944 & 0.334 & 0.652\\
(iv) Mixed shape
& 0.638 & 0.696 & 0.188 & 0.374
& 0.960 & 0.982 & 0.430 & 0.656\\
(v) Piecewise effect
& 0.282 & 0.502 & 0.130 & 0.238
& 0.630 & 0.810 & 0.200 & 0.494\\
(vi) Covariance-matched higher-order
& 0.664 & 0.146 & 0.352 & 0.102
& 0.926 & 0.106 & 0.544 & 0.128\\
\bottomrule
\end{tabular}
\end{table}

\subsection{Multiple treatments against control}\label{sec:simmulti}

We next consider the setting with three treatment groups and a shared control group. For size assessment, we use the eight null designs introduced in Section~\ref{sec:simdesign}. For the power experiments, let $Y_{ai}^0$ denote the untreated potential outcome of individual $i$ sampled in arm $a$, where $a=0$ denotes the control arm and $a=1,2,3$ the treatment arms. All $Y_{ai}^0$ are independent across individuals and arms and follow the Gaussian reference law in~\eqref{eq:simbase}. The observed control curves are $Y_{0i}^0$, while for treatment arm $a=1,2,3$ the observed curves are
\[
Y_{ai}^a(t)=Y_{ai}^0(t)+\tau_{ai}(t),
\]
where $\tau_{ai}$ is the individual treatment effect under treatment $a$. We keep the untreated-outcome law fixed and vary the treatment effects across the following five alternatives.
\begin{enumerate}[label=(M\arabic*),itemsep=0.45em]

\item \textit{Group locations with random individual effects.} Set $\tau_{ai}(t)=(a-1)+\eta_{ai}$, where $\eta_{ai}\stackrel{\mathrm{i.i.d.}}{\sim}N(0,1)$. The treatment groups therefore have different common locations together with individual-level variation.

\item \textit{Group-specific shapes.} Set $\tau_{ai}(t)=b_{ai}f_a(t)$, where $b_{ai}\stackrel{\mathrm{i.i.d.}}{\sim}N(0,1)$ and
\[
f_1(t)=\sin(2\pi t),\qquad
f_2(t)=t,\qquad
f_3(t)=2\sin(\pi t).
\]

\item \textit{Mixed shapes.} Let $b_{ai}\stackrel{\mathrm{i.i.d.}}{\sim}N(0,1)$ and $D_{ai}\sim\operatorname{Bernoulli}(p_a)$, where $p_a=(a-1)/2$, and set
\[
\tau_{ai}(t)
=
b_{ai}\left\{
D_{ai}t+(1-D_{ai})2\sin(\pi t)
\right\}.
\]
Thus the probabilities of the linear effect are $0$, $1/2$, and $1$ across the three treatment groups.

\item \textit{Smooth deviations from a common effect.} Set
\[
\tau_{ai}(t)=\sin(2\pi t)+\delta_a g_{ai}(t),
\]
where $\delta_a=(a-1)/2$ and $g_{ai}(t)=\sum_{r=1}^{5}\zeta_{air}\psi_r(t)$ with $\zeta_{air}\sim N(0,r^{-1})$. The basis functions are $\bigl(\psi_1,\ldots,\psi_5\bigr)=\sqrt{2}\bigl(\sin(2\pi t),\cos(2\pi t),\sin(4\pi t),\cos(4\pi t),\sin(6\pi t)\bigr)$. The first treatment group has only the common effect, while the second and third groups include additional individual variation.

\item \textit{Covariance-matched higher-order alternative.} Let $Q_{ai}=\sqrt{1-\theta}\,\xi_{ai1}+\sqrt{\theta}\,S_{ai}$, where $S_{ai}$ takes values $-1$ and $1$ with equal probability, and set
\[
Y_{ai}^a(t)
=
Q_{ai}e_1(t)
+
\sum_{j=2}^{100}j^{-0.55}\xi_{aij}e_j(t),
\qquad
\theta=0.95.
\]
Equivalently, $\tau_{ai}(t)=\{(\sqrt{1-\theta}-1)\xi_{ai1}+\sqrt{\theta}S_{ai}\}e_1(t)$. The treatment and control laws have the same mean and covariance operator, but their difference distributions are distinct. Indeed, $\phi_a^\Delta(ue_1)=e^{-(1-\theta)u^2}\cos^2(\sqrt{\theta}\,u)$ differs from $\phi_0^\Delta(ue_1)=e^{-u^2}$. This design therefore isolates higher-order heterogeneity not reflected in covariance differences.

\end{enumerate}

In designs (M1)--(M4), the random components of the treatment effects are independent of the baseline processes. Design (M5) instead allows the treatment effect to depend on the baseline first score.

\begin{table}[!htbp]
\centering
\caption{Null rejection frequencies for three treatments against a shared control}
\label{tab:multisize}
\scriptsize
\setlength{\tabcolsep}{3pt}
\begin{tabular}{lrrrrrrrr}
\toprule
& \multicolumn{4}{c}{$n=120$} & \multicolumn{4}{c}{$n=240$}\\
\cmidrule(lr){2-5}\cmidrule(lr){6-9}
Design
& Proposed & COVM & S-MMD & S-ED
& Proposed & COVM & S-MMD & S-ED\\
\midrule
(N1) Gaussian
& 0.044 & 0.080 & 0.052 & 0.066
& 0.068 & 0.078 & 0.046 & 0.054\\
(N2) Skewed
& 0.062 & 0.120 & 0.060 & 0.058
& 0.062 & 0.072 & 0.054 & 0.032\\
(N3) Student $t_3$
& 0.084 & 0.082 & 0.046 & 0.052
& 0.058 & 0.040 & 0.058 & 0.040\\
(N4) Discrete
& 0.066 & 0.110 & 0.052 & 0.056
& 0.068 & 0.076 & 0.054 & 0.056\\
(N5) Unequal allocation
& 0.054 & 0.118 & 0.050 & 0.068
& 0.068 & 0.092 & 0.068 & 0.050\\
(N6) Weak null
& 0.068 & 0.098 & 0.028 & 0.030
& 0.064 & 0.062 & 0.052 & 0.056\\
(N7) Near-degenerate
& 0.068 & 0.008 & 0.036 & 0.036
& 0.056 & 0.002 & 0.060 & 0.060\\
(N8) Cauchy
& 0.000 & --- & 0.044 & 0.046
& 0.004 & --- & 0.054 & 0.054\\
\bottomrule
\end{tabular}
\end{table}

Table~\ref{tab:multisize}  shows that the proposed test maintains reasonably stable size across a broad range of null distributions and sampling configurations. Its rejection frequencies remain close to the nominal level under skewed, heavy-tailed, discrete, unequal-allocation, and weak difference-law nulls, indicating robustness beyond the Gaussian setting. The near-degenerate design is also well behaved in the multiple-treatment setting, whereas the Cauchy design leads to pronounced conservativeness. In comparison, COVM exhibits noticeable size distortion in several settings and is not applicable when the population covariance is undefined.

\begin{table}[!htbp]
\centering
\caption{Alternative rejection frequencies for three treatments against a shared control}
\label{tab:multipower}
\scriptsize
\setlength{\tabcolsep}{3pt}
\begin{tabular}{lrrrrrrrr}
\toprule
& \multicolumn{4}{c}{$n=120$} & \multicolumn{4}{c}{$n=240$}\\
\cmidrule(lr){2-5}\cmidrule(lr){6-9}
Design
& Proposed & COVM & S-MMD & S-ED
& Proposed & COVM & S-MMD & S-ED\\
\midrule
(M1) Random individual effects
& 0.464 & 0.714 & 0.122 & 0.160
& 0.772 & 0.898 & 0.240 & 0.420\\
(M2) Group-specific shapes
& 0.472 & 0.880 & 0.178 & 0.252
& 0.906 & 0.992 & 0.270 & 0.636\\
(M3) Mixed shapes
& 0.560 & 0.912 & 0.170 & 0.268
& 0.914 & 0.994 & 0.348 & 0.664\\
(M4) Smooth deviations
& 0.498 & 0.750 & 0.152 & 0.380
& 0.956 & 0.974 & 0.424 & 0.852\\
(M5) Covariance-matched higher-order
& 0.690 & 0.120 & 0.116 & 0.058
& 1.000 & 0.092 & 0.238 & 0.052\\
\bottomrule
\end{tabular}
\end{table}

Table~\ref{tab:multipower} demonstrates that the proposed test gains substantial power as the sample size increases and consistently outperforms the two symmetry-preserving paired benchmarks across all five alternatives. COVM can attain higher rejection frequencies when heterogeneity directly changes the covariance structure, but the covariance-matched higher-order design highlights the broader sensitivity of the proposed procedure. When the treatment and control covariance operators are identical, the proposed rejection frequency increases from 0.690 to 1.000, whereas COVM remains close to its own null rejection level and both paired benchmarks are substantially less sensitive. These results show that the proposed test combines stable calibration across diverse null settings with strong sensitivity to distributional heterogeneity beyond second-order differences. Additional sensitivity analyses on numerical implementation and power under higher-order and local alternatives are not reproduced in this preprint version.

\FloatBarrier
\section{Real data analysis}\label{sec:realdata}
We apply the proposed test to the gas smart-metering Customer Behaviour Trial conducted by the Commission for Energy Regulation of Ireland. The trial examined how billing, consumption feedback, and pricing schemes affect residential gas use \citep{CER2011}. Our question is whether a given intervention produces a common change in households' consumption curves or whether its effect varies across households. The functional formulation allows the effect to vary over the course of a month, so homogeneity refers to agreement across households in the entire effect curve.

Gas consumption was recorded every half hour. The benchmark period ran from December 2009 to May 2010, followed by an experimental period from June 2010 to May 2011. The analysis sample contains 1,493 households assigned to one control group and four treatment groups, with group membership fixed throughout the study period. During the experimental period, the control group retained the standard tariff and billing every two months, while the four treatment groups received additional information, more frequent bills, an in-home display, or a seasonal tariff. The treatment definitions and corresponding group sizes are summarised in Table~\ref{tab:cer_groups}. We focus on the first seven intervention months, from June to December 2010, covering the initial summer period and the transition into winter. This window is used throughout the analysis. For each household, we construct a monthly consumption curve from the available half-hour readings. A small number of time points are jointly missing for all households in September and are omitted without imputation.  Further details on the monthly observation grids and data preparation are not reproduced in this preprint version.


\begin{table}[!htbp]
\centering
\caption{CER Gas interventions and analysis sample}
\label{tab:cer_groups}
\small
\begin{tabular}{lp{8.7cm}r}
\toprule
Group & Intervention & Customers\\
\midrule
Control & Standard tariff and a bill every two months & 524\\
Treatment 1 & Bill every two months and energy usage statement & 236\\
Treatment 2 & Monthly bill and energy usage statement & 228\\
Treatment 3 & Bill every two months, energy usage statement, and in-home display & 251\\
Treatment 4 & Treatment 3 package with a seasonal tariff & 254\\
\midrule
Total & & 1,493\\
\bottomrule
\end{tabular}
\end{table}

For a fixed month, let $Y_i^k(t)$ denote household $i$'s potential gas consumption at time $t$ under treatment $k$, and let $Y_i^0(t)$ denote its potential consumption under control. Here $t$ runs through the observed half-hours of that month. Only the curve corresponding to the household's assigned group is observed. We suppress the month index and apply the following formulation separately in each of the seven months. For $k=1,\ldots,4$, the single-treatment null hypothesis is
\[
H_0^k:\quad
Y_i^k(t)-Y_i^0(t)=\tau_k(t)
\quad\text{for all households }i\text{ and observed times }t,
\]
for some deterministic function $\tau_k(t)$, with the equality understood almost surely under the population model. Thus, under $H_0^k$, all households share the same treatment-effect curve, although their underlying consumption curves may differ. The alternative allows the treatment effect to vary across households.

We also consider the multiple-treatment hypothesis
\[
H_0^5:\quad
Y_i^k(t)-Y_i^0(t)=\tau_k(t),
\qquad k=1,\ldots,4,
\]
for all households and observed times, almost surely, where $\tau_1,\ldots,\tau_4$ are deterministic functions. Thus, each treatment may have its own common effect curve, and the alternative allows treatment-effect heterogeneity under at least one treatment. As shown in Section~\ref{sec:hypotheses}, treatment-effect homogeneity implies equality between the within-treatment and within-control difference distributions. We therefore test this restriction separately for each treatment and jointly across the four treatments using the shared control group, with the analysis repeated for each month.

We use the squared-modulus U-statistic with $M=256$ fixed Gaussian directions and $B=999$ multiplier draws. The curve representation retains all observed half-hours and uses covariance eigenvalues $\gamma_j=2j^{-1.2}$. Each household receives one multiplier for its entire curve, and  the same control multipliers are used across the four contrasts in the joint test. This accounts for dependence among a household's half-hour readings and for the shared control sample.

\begin{figure}[!htbp]
\centering
\includegraphics[width=\textwidth]{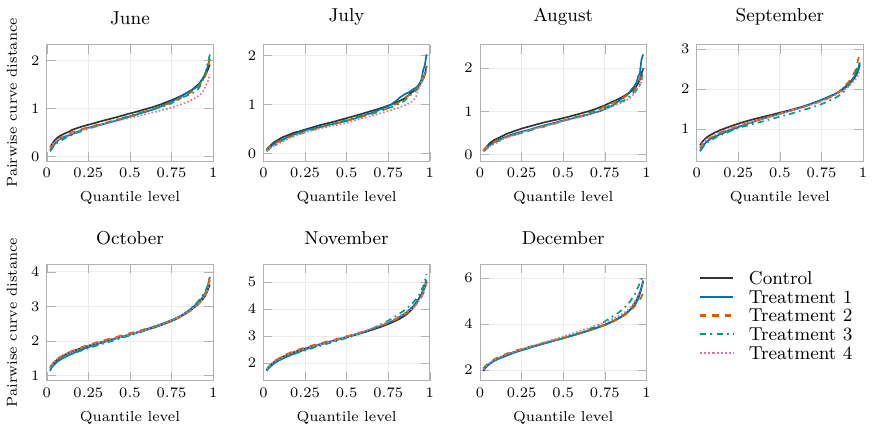}
\caption{Empirical quantile functions of within-arm pairwise curve distances, June--December 2010.}
\label{fig:cer_pairwise}
\end{figure}

For a descriptive view of the same restriction, Figure~\ref{fig:cer_pairwise} compares within-arm pairwise curve distances. For two households $i\ne j$ in arm $k$ during month $m$, define
\[
D_{ij,m}^{k}
=
\left\{
\frac{1}{q_m}\sum_{r=1}^{q_m}
\left(X_{imr}^{k}-X_{jmr}^{k}\right)^2
\right\}^{1/2}.
\]
Under $H_0^k$, the distribution of $D_{ij,m}^{k}$ agrees with that of $D_{ij,m}^{0}$. These distances summarize only the magnitude of the full functional differences and are used for descriptive purposes. Figure~\ref{fig:cer_pairwise} shows that the treatment and control quantile curves overlap closely over much of their ranges, particularly in the later months. The largest visual discrepancy occurs for Treatment 4 in June, where the treatment curve tends to lie below the control curve. Formal inference is based on the full difference distributions, which retain the direction and temporal structure of the functional differences.

The joint tests displayed in Table~\ref{tab:cer_pvalues}  provide the primary month-level assessment. Their $p$-values range from 0.440 to 0.891, providing no evidence that the observable restriction fails for at least one treatment in any month. The treatment-specific results show a similar pattern. The smallest unadjusted $p$-value is 0.046 for Treatment 4 in June, while all remaining treatment-specific $p$-values are at least 0.181. The June Treatment 4 result is an isolated unadjusted signal. It is not repeated in subsequent months, and the corresponding June joint test gives $p=0.440$. Moreover, Treatment 4 combines several intervention components, so this comparison does not identify a particular mechanism underlying the observed discrepancy. Across the seven months, we report five tests per month, giving 35 hypotheses in total. Because the same households are followed over time and all treatment comparisons share the control group, the resulting tests are dependent. Holm's procedure controls the familywise error rate under such dependence. After adjustment across the 35 tests, no comparison rejects at the 5\% familywise level, with all adjusted $p$-values equal to 1.000.

\begin{table}[!htbp]
\centering
\caption{Monthly multiplier-bootstrap $p$-values for the CER Gas data}
\label{tab:cer_pvalues}
\small
\begin{tabular}{lrrrrr}
\toprule
Month & Treatment 1 & Treatment 2 & Treatment 3 & Treatment 4 & Joint\\
\midrule
June      & 0.712 & 0.465 & 0.939 & 0.046 & 0.440\\
July      & 0.993 & 0.310 & 0.807 & 0.390 & 0.779\\
August    & 0.942 & 0.310 & 0.911 & 0.627 & 0.891\\
September & 0.364 & 0.921 & 0.862 & 0.523 & 0.850\\
October   & 0.327 & 0.685 & 0.685 & 0.557 & 0.748\\
November  & 0.185 & 0.388 & 0.959 & 0.339 & 0.462\\
December  & 0.181 & 0.889 & 0.863 & 0.655 & 0.725\\
\bottomrule
\end{tabular}
\end{table}

Overall, we find no multiplicity-adjusted evidence against the observable difference-distribution restrictions implied by treatment-effect homogeneity during June to  December 2010. The same qualitative conclusion is supported by the pre-intervention diagnostic and remains stable across the numerical sensitivity analyses not reproduced in this preprint version. These nonrejections do not establish individual-level homogeneity, and weaker or less detectable forms of heterogeneity may still be present. The empirical-reference experiments not reproduced in this preprint version further illustrate how detectability depends on the magnitude and structure of heterogeneous effects.

\section{Discussion}
\label{sec:discussion}
This paper develops a nonparametric framework for assessing treatment-effect homogeneity with functional outcomes through a testable marginal implication. The proposed discrepancy is invariant to an unknown common functional effect and, with a full-support Gaussian integrating measure, characterizes equality of the corresponding within-group difference distributions. We derive the asymptotic behavior of the resulting U-statistic under the null and alternatives, develop a centered multiplier bootstrap for inference, and provide a computational implementation based on random directions and exact pairwise aggregation. Simulations demonstrate reliable calibration and sensitivity to a range of heterogeneous alternatives, while the Irish gas smart-metering application illustrates how the method can be used to examine heterogeneity in functional responses.

Several limitations also suggest directions for future work. The procedure tests a marginal implication of individual treatment-effect homogeneity and therefore cannot detect heterogeneous effects that leave the relevant difference distributions unchanged. In addition, the current framework considers a fixed treatment assignment over time. An important extension would allow treatments or treatment regimes themselves to vary over time, which is common in longitudinal studies. Further work could also accommodate more complex treatment-assignment mechanisms and dependence structures.

\appendix
\FloatBarrier
\section{Proof outlines}
\label{app:main-proofs}

The arguments below summarize the main steps behind the theoretical results. Detailed combinatorial calculations for the generalized U-statistics, moment estimates for higher-order kernels, and trace-class approximation arguments are not reproduced in this preprint version.

\paragraph{Identification (Proposition~\ref{prop:ident}).}
If the independent-copy difference laws coincide, their characteristic functionals coincide, and hence $T=0$. Conversely, if the laws differ, uniqueness of characteristic functionals gives a direction $z_0$ at which $\Delta_\phi(z_0)\ne0$. Continuity makes $|\Delta_\phi|$ bounded away from zero on a neighborhood of $z_0$. Injectivity of $\GZ$ gives full support of the Gaussian integrating measure, so this neighborhood has positive measure and $T>0$.

\paragraph{Projection expansion (Proposition~\ref{prop:expansion}).}
Decompose the pair feature into its expectation, first-order projection, and canonical second-order component. The corresponding pair average obeys~\eqref{eq:pairaverage}. Applying this decomposition to both features in the within-group fourth-order U-statistics, and separately to the between-group product, yields~\eqref{eq:expansion} by collecting population, linear, quadratic, and higher-order terms. The canonical degeneracy and boundedness of the feature imply, under the sampling proportions in the paper,
\[
\E\|U_{d,n}^{(2)}\|^2=O(n_d^{-2}),\qquad
\E Q_n^2=O(n^{-2}),\qquad \E R_n^2=O(n^{-3}).
\]
These orders hold also for independent, rowwise-changing outcome laws. The exact kernel-by-kernel calculations are omitted here.

\paragraph{Null limit (Theorem~\ref{thm:null}).}
Under the observable null, $\Delta_\phi=0$, and the projection expansion gives $n\TU=nQ_n+o_{\Prob}(1)$. Removing diagonal terms from the quadratic U-statistic produces
\[
 nQ_n=\|G_n\|^2-\tr(C_n)+o_{\Prob}(1),\qquad
 G_n=\sqrt n(\bar\psi_1-\bar\psi_0),\quad C_n=\sum_{d=0}^1\frac{n}{n_d}C_d.
\]
The bounded Hilbert-valued projections satisfy finite-dimensional Lindeberg conditions. The trace-class covariance tail bounds extend their finite-dimensional convergence to $G_n\Rightarrow G$ in $L^2(\HH,\mu)$. The quadratic continuous mapping and trace convergence yield the stated centered Gaussian quadratic-form limit.

\paragraph{Fixed and local alternatives (Theorems~\ref{thm:fixed} and~\ref{thm:local}).}
Under fixed laws, the higher-order terms contribute $O_{\Prob}(n^{-1})$, so
\[
 \TU-T=2\langle\Delta_\phi,\bar\psi_1-\bar\psi_0\rangle+O_{\Prob}(n^{-1}).
\]
A scalar central limit theorem establishes the normal limit when its variance is positive. Under local row laws, the uniform moment bounds instead yield
\[
 n\TU=\|G_n+\sqrt n\,\Delta_{\phi,n}\|^2-\tr(C_n)+o_{\Prob}(1).
\]
The bounded triangular-array projections and the assumed trace-norm covariance convergence give $G_n\Rightarrow G$, proving the shifted quadratic limit.

\paragraph{Multiplier bootstrap (Theorem~\ref{thm:bootstrap}).}
The empirical projections in~\eqref{eq:ahat} approximate their population counterparts in average mean square at rate $O(n_d^{-1})$. Under the stated conditions, their aggregate empirical covariance therefore converges in trace norm to $C$. Conditional on the observations, the multiplier summands are independent, centered and satisfy the finite-dimensional Lindeberg condition. Covariance-tail control extends the conditional limit to $G_n^*\Rightarrow G$ in probability, and the centered quadratic mapping gives the limit of~\eqref{eq:bootstrap}. If $C\ne0$, the limiting quadratic form has a continuous distribution at its upper quantile. The null-level and local-power claims follow from quantile convergence; under fixed alternatives, the statistic diverges while the conditional critical value remains bounded in probability.

\paragraph{Multiple treatment groups (Theorem~\ref{thm:multi}).}
Collect the $m$ treatment--control comparisons into a product Hilbert space. Because the same control sample appears in every comparison, the covariance block for coordinates $d$ and $\ell$ is
\[
 (\boldsymbol C_{m,n})_{d\ell}
 =\ind\{d=\ell\}\frac n{n_d}C_d+\frac n{n_0}C_0.
\]
The fixed number of arms permits the binary projection estimates and finite-dimensional approximation arguments to be applied in this product space. Reusing each control multiplier across the comparisons preserves the off-diagonal covariance blocks and yields the corresponding conditional bootstrap limit. The fixed- and local-alternative extensions follow from the same linear and shifted-quadratic expansions.

\FloatBarrier
\begingroup
\small
\raggedright

\endgroup

\begin{thebibliography}{99}
\bibitem[Bogachev(1998)]{Bogachev1998}
Bogachev, V.~I. (1998).
\textit{Gaussian Measures}. Mathematical Surveys and Monographs, Vol.~62.
American Mathematical Society, Providence, RI.

\bibitem[Bugni and Horowitz(2021)]{Bugni2021}
Bugni, F.~A. and Horowitz, J.~L. (2021).
Permutation tests for equality of distributions of functional data.
\textit{Journal of Applied Econometrics}, 36(7), 861--877.
\href{https://doi.org/10.1002/jae.2846}{doi:10.1002/jae.2846}.

\bibitem[Chung and Olivares(2021)]{Chung2021}
Chung, E. and Olivares, M. (2021).
Permutation test for heterogeneous treatment effects with a nuisance parameter.
\textit{Journal of Econometrics}, 225(2), 148--174.
\href{https://doi.org/10.1016/j.jeconom.2020.09.015}{doi:10.1016/j.jeconom.2020.09.015}.

\bibitem[Commission for Energy Regulation(2011)]{CER2011}
Commission for Energy Regulation (2011).
\textit{Gas Customer Behaviour Trial Findings Report}.
Smart Metering Information Paper CER11180a, 11 October 2011.

\bibitem[Crump et~al.(2008)]{crump2008nonparametric}
Crump, R.~K., Hotz, V.~J., Imbens, G.~W., and Mitnik, O.~A. (2008).
Nonparametric tests for treatment effect heterogeneity.
\textit{The Review of Economics and Statistics}, 90(3), 389--405.
\href{https://doi.org/10.1162/rest.90.3.389}{doi:10.1162/rest.90.3.389}.

\bibitem[Cuevas et~al.(2004)]{cuevas2004anova}
Cuevas, A., Febrero, M., and Fraiman, R. (2004).
An ANOVA test for functional data.
\textit{Computational Statistics \& Data Analysis}, 47(1), 111--122.

\bibitem[Ding et~al.(2016)]{Ding2016}
Ding, P., Feller, A., and Miratrix, L. (2016).
Randomization inference for treatment effect variation.
\textit{Journal of the Royal Statistical Society, Series B}, 78(3), 655--671.
\href{https://doi.org/10.1111/rssb.12124}{doi:10.1111/rssb.12124}.

\bibitem[Ecker et~al.(2024)]{ecker2024causal}
Ecker, K., de Luna, X., and Schelin, L. (2024).
Causal inference with a functional outcome.
\textit{Journal of the Royal Statistical Society, Series C}, 73(1), 221--240.
\href{https://doi.org/10.1093/jrsssc/qlad092}{doi:10.1093/jrsssc/qlad092}.

\bibitem[Gretton et~al.(2012)]{Gretton2012}
Gretton, A., Borgwardt, K.~M., Rasch, M.~J., Sch\"olkopf, B., and Smola, A. (2012).
A kernel two-sample test.
\textit{Journal of Machine Learning Research}, 13(25), 723--773.
\url{https://jmlr.org/papers/v13/gretton12a.html}.

\bibitem[Hoeffding(1948)]{Hoeffding1948}
Hoeffding, W. (1948).
A class of statistics with asymptotically normal distribution.
\textit{Annals of Mathematical Statistics}, 19(3), 293--325.
\href{https://doi.org/10.1214/aoms/1177730196}{doi:10.1214/aoms/1177730196}.

\bibitem[Hsu and Shen(2019)]{hsu2019testing}
Hsu, Y.-C. and Shen, S. (2019).
Testing treatment effect heterogeneity in regression discontinuity designs.
\textit{Journal of Econometrics}, 208(2), 468--486.

\bibitem[Imai and Ratkovic(2013)]{imai2013estimating}
Imai, K. and Ratkovic, M. (2013).
Estimating treatment effect heterogeneity in randomized program evaluation.
\textit{The Annals of Applied Statistics}, 7(1), 443--470.
\href{https://doi.org/10.1214/12-AOAS593}{doi:10.1214/12-AOAS593}.

\bibitem[Lyons(2013)]{Lyons2013}
Lyons, R. (2013).
Distance covariance in metric spaces.
\textit{Annals of Probability}, 41(5), 3284--3305.
\href{https://doi.org/10.1214/12-AOP803}{doi:10.1214/12-AOP803}.

\bibitem[Paparoditis and Sapatinas(2016)]{Paparoditis2016}
Paparoditis, E. and Sapatinas, T. (2016).
Bootstrap-based testing of equality of mean functions or equality of covariance operators for functional data.
\textit{Biometrika}, 103(3), 727--733.
\href{https://doi.org/10.1093/biomet/asw033}{doi:10.1093/biomet/asw033}.

\bibitem[Rahimi and Recht(2007)]{Rahimi2007}
Rahimi, A. and Recht, B. (2007).
Random features for large-scale kernel machines.
In \textit{Advances in Neural Information Processing Systems 20}, pp.~1177--1184.

\bibitem[Ramirez-Cuellar(2023)]{Ramirez2023}
Ramirez-Cuellar, J. (2023).
Testing for idiosyncratic treatment effect heterogeneity.
\textit{arXiv preprint}, arXiv:2304.01141.
\url{https://arxiv.org/abs/2304.01141}.

\bibitem[Ramsay and Silverman(2005)]{RS05}
Ramsay, J.~O. and Silverman, B.~W. (2005).
\textit{Functional Data Analysis}, 2nd edition.
Springer, New York.

\bibitem[The Research Perspective Ltd(2012)]{ResearchPerspective2012}
The Research Perspective Ltd (2012).
\textit{Manifest: Smart Meter Gas Trial Data}.
Data documentation, 15 October 2012.

\bibitem[Sant'Anna(2021)]{SantAnna2021}
Sant'Anna, P.~H.~C. (2021).
Nonparametric tests for treatment effect heterogeneity with duration outcomes.
\textit{Journal of Business \& Economic Statistics}, 39(3), 816--832.

\bibitem[Serfling(1980)]{Serfling1980}
Serfling, R.~J. (1980).
\textit{Approximation Theorems of Mathematical Statistics}.
Wiley, New York.

\bibitem[Sejdinovic et~al.(2013)]{Sejdinovic2013}
Sejdinovic, D., Sriperumbudur, B.~K., Gretton, A., and Fukumizu, K. (2013).
Equivalence of distance-based and RKHS-based statistics in hypothesis testing.
\textit{The Annals of Statistics}, 41(5), 2263--2291.
\href{https://doi.org/10.1214/13-AOS1140}{doi:10.1214/13-AOS1140}.




\bibitem[Testa et~al.(2026)]{testa2026doubly}
Testa, L., Boschi, T., Chiaromonte, F., Kennedy, E.~H., and Reimherr, M. (2026).
Doubly-robust functional average treatment effect estimation.
\textit{Journal of Causal Inference}, 14(1), 20250045.
\href{https://doi.org/10.1515/jci-2025-0045}{doi:10.1515/jci-2025-0045}.

\bibitem[Wang et~al.(2016)]{Wang-review2016}
Wang, J.-L., Chiou, J.-M., and M\"uller, H.-G. (2016).
Functional data analysis.
\textit{Annual Review of Statistics and Its Application}, 3, 257--295.
\href{https://doi.org/10.1146/annurev-statistics-041715-033624}{doi:10.1146/annurev-statistics-041715-033624}.

\bibitem[Wynne and Duncan(2022)]{Wynne2022}
Wynne, G. and Duncan, A.~B. (2022).
A kernel two-sample test for functional data.
\textit{Journal of Machine Learning Research}, 23(73), 1--51.
\url{https://jmlr.org/papers/v23/20-1180.html}.


\bibitem[Zhang et~al.(2025)]{Zhang2025}
Zhang, Y., Wang, R., and Shao, X. (2025).
Adaptive testing for high-dimensional data.
\textit{Journal of the American Statistical Association}, 120(551), 1893--1905.
\href{https://doi.org/10.1080/01621459.2024.2439617}{doi:10.1080/01621459.2024.2439617}.



\end{thebibliography}
\end{document}